\documentclass[12pt]{iopart}

\usepackage{graphicx}
\usepackage{subfig}
\usepackage{graphbox}
\usepackage[dvipsnames]{xcolor}
\def\be{\begin{equation}}
\def\ee{\end{equation}}
\def\bea{\begin{eqnarray}}
\def\eea{\end{eqnarray}}
\newcommand{\ket}[1]{\mbox{$|#1\rangle$}}
\newcommand{\bra}[1]{\mbox{$\langle#1|$}}
\newcommand{\avg}[1]{\mbox{$\langle#1\rangle$}}

\begin{document}

\title{Transient dynamics of the quantum light retrieved from Rydberg polaritons}

\author{Auxiliadora Padr\'on-Brito$^1$, Roberto Tricarico$^1$, Pau Farrera $^{1,2}$, Emanuele Distante $^{1,2}$, Klara Theophilo$^1$, Darrick Chang$^{1,3}$ and Hugues de Riedmatten $^{1,3}$}

\address{$^1$ ICFO-Institut de Ciencies Fotoniques, The Barcelona Institute of Science and Technology, 08860 Castelldefels, Barcelona, Spain}
\address{$^2$ Max-Planck-Institut f{\"u}r Quantenoptik, Garching, Germany}
\address{$^3$ ICREA-Instituci{\'o} Catalana de Recerca i Estudis Ava\c{c}ats, 08015 Barcelona, Spain}

\ead{auxiliadora.padron@icfo.eu}

\begin{abstract}
We study the photon statistics of weak coherent pulses propagating through a cold Rydberg atomic ensemble in the regime of Rydberg electromagnetically induced transparency. We show experimentally that the value of the second-order autocorrelation function of the transmitted light strongly depends on the position within the pulse and heavily varies during the transients of the pulse. In particular, we show that the falling edge of the transmitted pulse displays much lower values than the rest of the pulse. We derive a theoretical model that quantitatively predicts our results and explains the physical behavior involved. Finally, we use this effect to generate single photons localized within a pulse from the atomic ensemble. We show that by selecting only the last part of the transmitted pulse, the single photons show an antibunching parameter as low as 0.12 and a generation efficiency per trial larger than possible with probabilistic generation schemes with atomic ensembles. 
\end{abstract}

\section{Introduction}
Highly excited Rydberg atoms enable long-range controllable atom-atom interactions \cite{Saffman2010}. This property has been used to demonstrate entanglement \cite{Wilk2010, Levine2018} and quantum gates between individual atoms \cite{Isenhower2010,Maller2015} and more recently quantum simulation in atomic arrays \cite{Bernien2017,Barredo2018}. Besides using individual atoms, another approach currently investigated is to use ensembles of highly excited Rydberg atoms, which on top of providing a medium with long-range controllable interaction between atoms also allows an efficient quantum light-matter interface. These properties lead to optical nonlinearities at the single-photon level \cite{Peyronel2012,Firstenberg2013,ParisMandoki2017} that can be used for quantum nonlinear optics \cite{Chang2014a,Firstenberg2016} and quantum information applications. The use of ensemble-based Rydberg mediated interactions has been explored within a range of schemes, such as single-photon level transistors and switches \cite{Tiarks2014a, Gorniaczyk2014a, Baur2014a}, photon-photon interactions \cite{Busche2017} and gates \cite{Tiarks2019}. Ensembles of Rydberg atoms can also be used to generate single photons in a quasi-deterministic fashion \cite{Dudin2012,Maxwell2013,Li2016,Li2016a}, contrary to ground state ensembles which usually rely on probabilistic or heralded schemes \cite{Duan2001}.

The use of electromagnetically induced transparency (EIT) with Rydberg excited levels \cite{Mohapatra2007} provides an efficient way of mapping light onto Rydberg excitations, leading to the creation of strong optical nonlinearities in the transmitted light thanks to the phenomenon of Rydberg blockade. Under EIT conditions, probe photons ($\Omega_p$) resonant with a ground-to-excited state transition $\ket{g}\to \ket{e}$ can coherently propagate within a narrow transparency window with the use of an additional resonant control field ($\Omega_c$) that couples $\ket{e}$ to a metastable state $\ket{r}$ (see Fig. 1c). This allows probe photons to be mapped onto dark-state polaritons, propagating with a group velocity $v_{gr}$ much smaller than the velocity of light in vacuum $c$. However, if $\ket{r}$ is a Rydberg level, the dipole-dipole interaction prevents the excitation of two atoms to the Rydberg state if they are closer than a distance called the blockade radius ($r_b$) \cite{Saffman2010}, which destroys the EIT transparency for two propagating probe photons within the distance $r_b$. As a consequence, strong scattering and absorption associated with a resonant two-level medium occurs, whereby the probability of detecting two photons at the same time in the transmitted light mode is ideally reduced to zero. In practice, within $r_b$, the amount of light transmitted in the two-level regime depends on the optical depth per blockade radius $D_b$, as $\exp(-D_{b})$.

The transmission of continuous-wave weak coherent input states of light through an ensemble of atoms under Rydberg EIT has been investigated theoretically \cite{Peyronel2012,Petrosyan2011,Gorshkov2013,Moos2015,Zeuthen2017}, and experimentally with the demonstration of single-photon nonlinearity and strong antibunching of the output light \cite{Peyronel2012,ParisMandoki2017}. However, the use of CW light renders challenging the efficient creation of single photons localized in time, as required for some applications in quantum information science \cite{sangouard2012single,Sangouard2007,knill2001}. It has been shown that localized single photons can be retrieved from collective Rydberg excitations, which can be created by using an off-resonant two-photon excitation scheme \cite{Dudin2012, Li2016, Li2016a,Craddock2019} or by storage of weak coherent light pulses under EIT conditions \cite{Maxwell2013,Maxwell2014}.
%It has been shown that single localized photons can be created by using an off-resonant two-photon excitation scheme to Rydberg states \cite{Dudin2012, Li2016, Li2016a,Craddock2019} or by storing weak coherent light pulses as collective Rydberg excitations by switching off the control laser in Rydberg EIT experiments \cite{Maxwell2013,Maxwell2014}.
However, this represents an additional experimental complexity and source of inefficiency. The use of transmitted pulses (without storage) in Rydberg EIT could therefore potentially lead to the generation of localized single photons with higher efficiencies. Nevertheless, the quantum statistics of weak pulses traveling under Rydberg EIT conditions has not been well studied yet. Moreover, as shown in \cite{Dudin2012} reaching low values of $g^{(2)}(0)$ requires very high Rydberg levels and high values of $D_b$, which are experimentally challenging to achieve, and low values of $g^{(2)}(\tau)$ are only observed for a small value of $\tau$ ($\sim$100 ns), such that reaching low values of $g^{(2)}$ for the entire pulse is challenging.

In this paper, we study experimentally and theoretically the propagation of weak coherent input pulses through a Rydberg EIT window. We measure the second-order correlation function $g^{(2)}_{\Delta t}$ for short time windows $\Delta t$ inside the pulse for various input pulse shapes. We show that the values of $g^{(2)}_{\Delta t}$ strongly vary during the transient phases, leading to very low values for the trailing edge, while the value for the full output pulse is much higher. Recently, similar behavior of $g^{(2)}$ during the EIT transients have been observed \cite{Moehl2020}. Here, we perform a systematic experimental and theoretical investigation to better understand the underlying physics. We develop a theoretical framework that can quantitatively predict our experimental results and explain the physical behavior involved. In addition, we investigate the use of the transient regime for the production of higher quality narrowband single photons by measuring $g^{(2)}_{\Delta t}$ as a function of the photon generation efficiency and comparing the results with the single photons obtained via storage in the Rydberg states.

\section{Experiment}\label{sec:Experiment}
\begin{figure}[h] 
	\centering
	\includegraphics[width=0.9\textwidth]{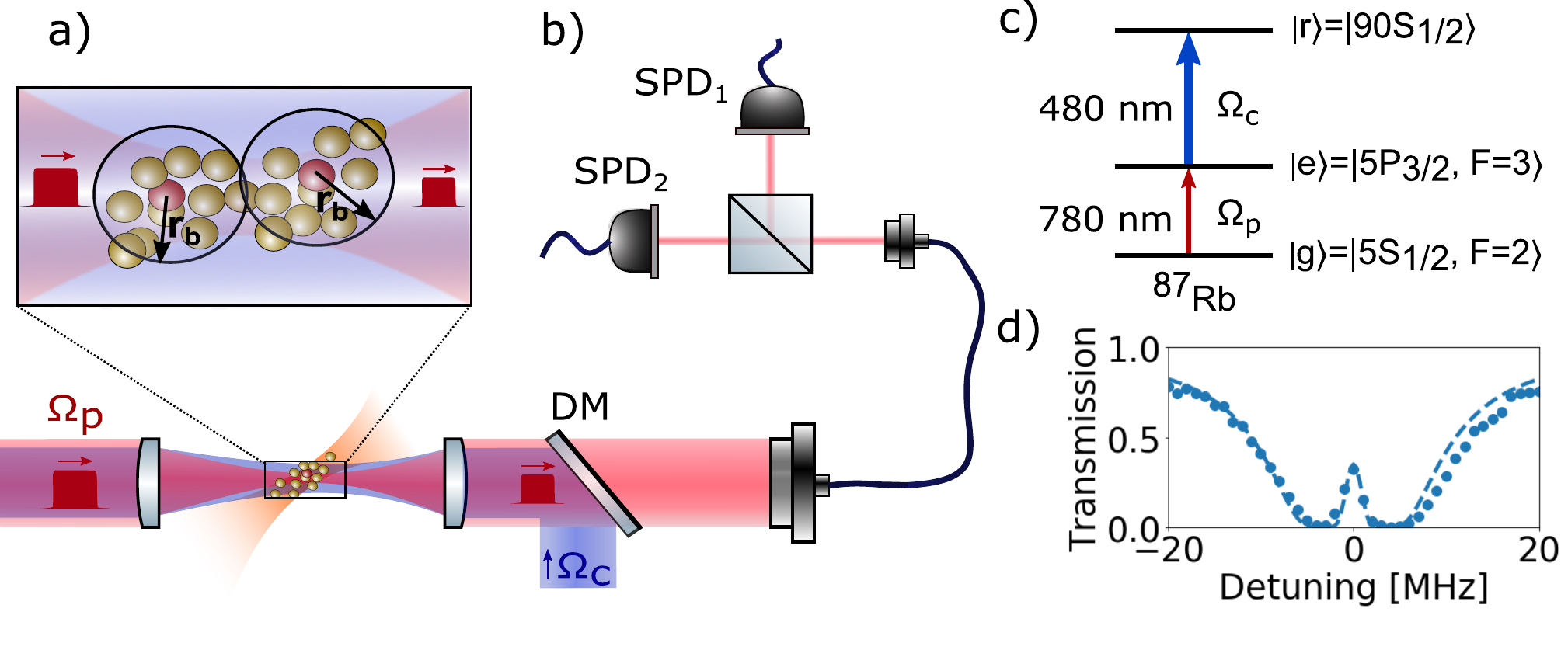}
	\caption{ (a) Schematic representation of the experimental setup.  An input probe pulse (in red) is sent to the experiment with a spatial distribution given by the probe beam (also in red), at a 19$^o$ angle to the dipole trap. The control beam (in blue) is counter-propagated with the probe and both are focused in the center of a dipole trap by two aspheric lenses. (b) The transmitted photons are sent through a Hanbury Brown and Twiss setup consisting of a beam-splitter and two single-photon detectors SPD$_1$ and SPD$_2$ to measure the second-order autocorrelation function. $\Omega_p$: Probe Rabi frequency, $\Omega_c$: Control Rabi frequency, DM: Dichroic mirror. (c) Level scheme and transitions used in the experiment. (d) EIT transparency window measurement (points) and theoretical simulation (dashed line) presented in Sec. \ref{theory}.}
	\label{fig:setup}
\end{figure}

Fig. \ref{fig:setup} shows a scheme of the experimental setup. A cloud of cold $^{87}$Rb atoms is loaded in a dipole trap, with an atomic peak density of $\sim 4\cdot10^{11}$ cm$^{-3}$ and a transversal size of 34 $\mu$m. The experimental sequence is the following: we start by loading a magneto-optical trap (MOT) for 2 seconds, followed by a compression of the MOT and a molasses period; the dipole trap is switched on 500 ms after the beginning of the MOT and it is switched off during the excitation of the atoms to the Rydberg level, to avoid losing atoms and remove AC Stark shifts; the dipole trap is modulated with a period of 16 $\mu$s, which leaves a time of less than 8 $\mu$s to perform one experimental trial. This is repeated 13000 times during the dipole trap lifetime, which gives a total experimental rate of 5.73 kHz.

The atoms are initially prepared in the ground state $|g\rangle=|5S_{1/2}, F=2\rangle$ and resonantly coupled with the state $|e\rangle=|5P_{3/2},F'=3\rangle$ by a weak probe field of 780 nm. The excited state is also resonantly coupled with the Rydberg state $|r\rangle=|90S_{1/2}\rangle$ using a counter-propagating control beam at 479.4 nm (see Fig. \ref{fig:setup}.c). The probe beam is sent with an angle of $19^o$ with respect to the dipole trap beam (see Fig. \ref{fig:setup}.a) and propagates through the medium with an optical depth of $D\approx10$. It is focused in the center of the atomic medium with a beam waist of $w_p\approx6.5 \ \mu$m. As the blockade radius is $r_b\approx13 \ \mu$m for our experimental parameters, the parallel propagation of two polaritons in our atomic cloud is negligible and the optical depth per blockade radius is $D_b\approx1$. Fig. \ref{fig:setup}.d shows an example of the transmission spectrum of Rydberg EIT in our cloud, for a weak single-photon level probe. By fixing the spontaneous decay rate of the excited level $\Gamma$ to be $\approx2\pi\times$6 MHz \cite{steck2001rubidium}, it results a control Rabi frequency of $2\Omega_c\approx2\pi\times$6.4 MHz, a dephasing rate of the Rydberg level of $\gamma_r\approx2\pi\times$0.8 MHz, and a full width at half maximum of the transmission (EIT bandwidth) of $\delta_{EIT}\approx2\pi\times2.3$ MHz.

The statistics of the light is obtained from the second-order autocorrelation function, defined as:
\begin{equation}\label{g2equation}
 g^{(2)}(t,\tau)=\frac{\left< E^{\dagger}(t)E^{\dagger}(t+\tau)E(t+\tau)E(t) \right>}{\left< E^{\dagger}(t)E(t) \right>\left< E^{\dagger}(t+\tau)E(t+\tau) \right>} \mathrm{,}
\end{equation}
where $E(t)$ is the electric field operator for the transmitted light mode at time t. We measure the second-order autocorrelation function by sending the output pulses through a Hanbury Brown and Twiss setup, which consists of a beam splitter and two single-photon avalanche detectors (SPDs), as shown in Fig. \ref{fig:setup}.b. The arrival times of the photons for each detector, together with trigger times for each trial, are saved in a time-stamp file.

First, we select a detection time window $\Delta t$. Then, $g^{(2)}_{\Delta t}(t,\tau)$ is obtained from the equation
\begin{equation}
g^{(2)}_{\Delta t}(t,\tau)=\frac{P_C (\Delta t)}{P_1 (\Delta t)P_2 (\Delta t)} \mathrm{,}
\end{equation}
where $t$ is the starting time of the detection window in the detector SPD$_1$ and $t+\tau$ is the starting time in the detector SPD$_2$, both measured with respect to the trigger time. $P_C (\Delta t)$ is the coincidence probability between the two detectors and $P_1 (\Delta t)$ and $P_2 (\Delta t)$ are the detection probabilities in detector SPD$_1$ and SPD$_2$, respectively. The normalization factor $P_1 (\Delta t)P_2 (\Delta t)$ is obtained by averaging the number of coincidences between photons arriving to the first detector in one trial and the photons arriving to the second detector from the 5th to the 20th following trials, where there is no correlation (see Fig. \ref{fig:g2ex}a). The correlation between the first five trials can be explained by the creation of long-lived pollutants \cite{Bienias2020,ornelas2020demand}. The transmission efficiency from the ensemble to the first detector is $0.23\pm0.02$, taking into account all the optical elements, and the SPD$_1$ detection efficiency is $0.43\pm0.04$.

\section{Time-resolved photon correlation}
\begin{figure}[h]
	\centering
	\includegraphics[width=0.9\textwidth]{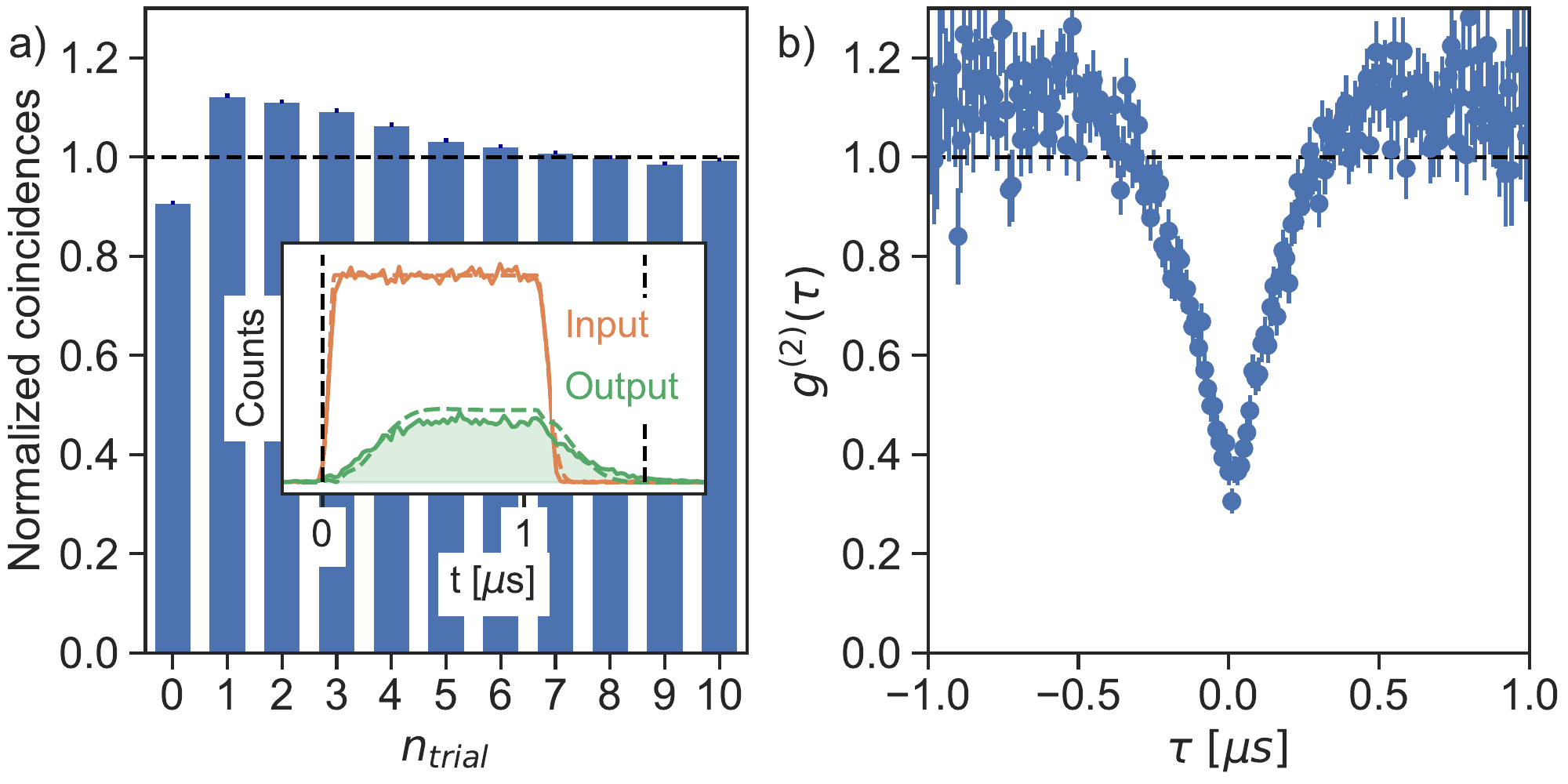}
	\caption{Statistics for a square pulse. (a) Normalized coincidences between the trial 0 and trial $n_{trial}$. Then, the first value, corresponding to $n_{trial}=0$, shows the normalized coincidences in the same trial, which gives a value of $g^{(2)}_{\Delta t}(0,0)=0.908\pm0.004$. Inset in (a) shows the temporal distribution of the input (in orange) and the output pulse (in green). The solid lines indicate the experimental values and the dashed lines the results of the theoretical simulations presented in Sec. \ref{theory}. The black lines delimit the temporal window with $\Delta t=1.6 \ \mu s$, longer that the output pulse duration. (b) Normalized coincidences as a function of the delay time $\tau$ between the two-photon events, being the one at zero-delay time $g^{(2)}(0)=0.31\pm0.03$. These values are calculated with a bin size of 10 ns. Error bars in the plots correspond to one standard deviation.} 
	\label{fig:g2ex}
\end{figure}

First, we study the photon statistics for the case of a square pulse with a temporal length of $1 \ \mu$s and a mean number of photons of $\approx1.5$, propagating through the Rydberg medium under EIT conditions. In the inset of Fig. \ref{fig:g2ex}.a, we can see the temporal shape of the input and output pulses, showing an EIT transmission efficiency of $\eta=N_{out}/N_{in} =0.285\pm0.016$, where $N_{out(in)}$ is the total number of counts in the output(input) pulse. Fig. \ref{fig:g2ex}.a, shows an example of the autocorrelation measurement for $\Delta t=1.6 \ \mu$s, sufficiently large to include the whole output pulse (see the detection window delimited by the black dashed lines of the inset). The normalized coincidences for the $n_{trial}=0$ leads to  $g^{(2)}_{\Delta t}(0,0) = 0.908\pm0.004$, which shows that while the full output pulse displays quantum statistics, it is still far from being a single photon. 

We then measure the time-resolved autocorrelation function $g^{(2)}(\tau)$,  where  $\tau$ is the delay time between two-photon detections within the whole output pulse. For that purpose, we measure the coincidences for different $\tau$, inside a bin size of 10 ns, and normalize them by the coincidences between the following 5th-20th trials. The results are shown in Fig. \ref{fig:g2ex}.b. For $\tau =0$, we obtain $g^{(2)}(0)=0.31 \pm 0.03$, demonstrating the single-photon nature of the output light, as shown previously in \cite{Peyronel2012,Maxwell2013,Gorniaczyk2014a,Tiarks2014a,Baur2014a}. However, we see that $g^{(2)} (\tau)$ quickly increases to 1 for $\tau \geq 350$ ns, much shorter than the pulse duration. This increase is attributed to the fact that the compressed pulse is longer than the blockade radius.

\begin{figure}[h]
	\centering
	\includegraphics[width=0.9\textwidth]{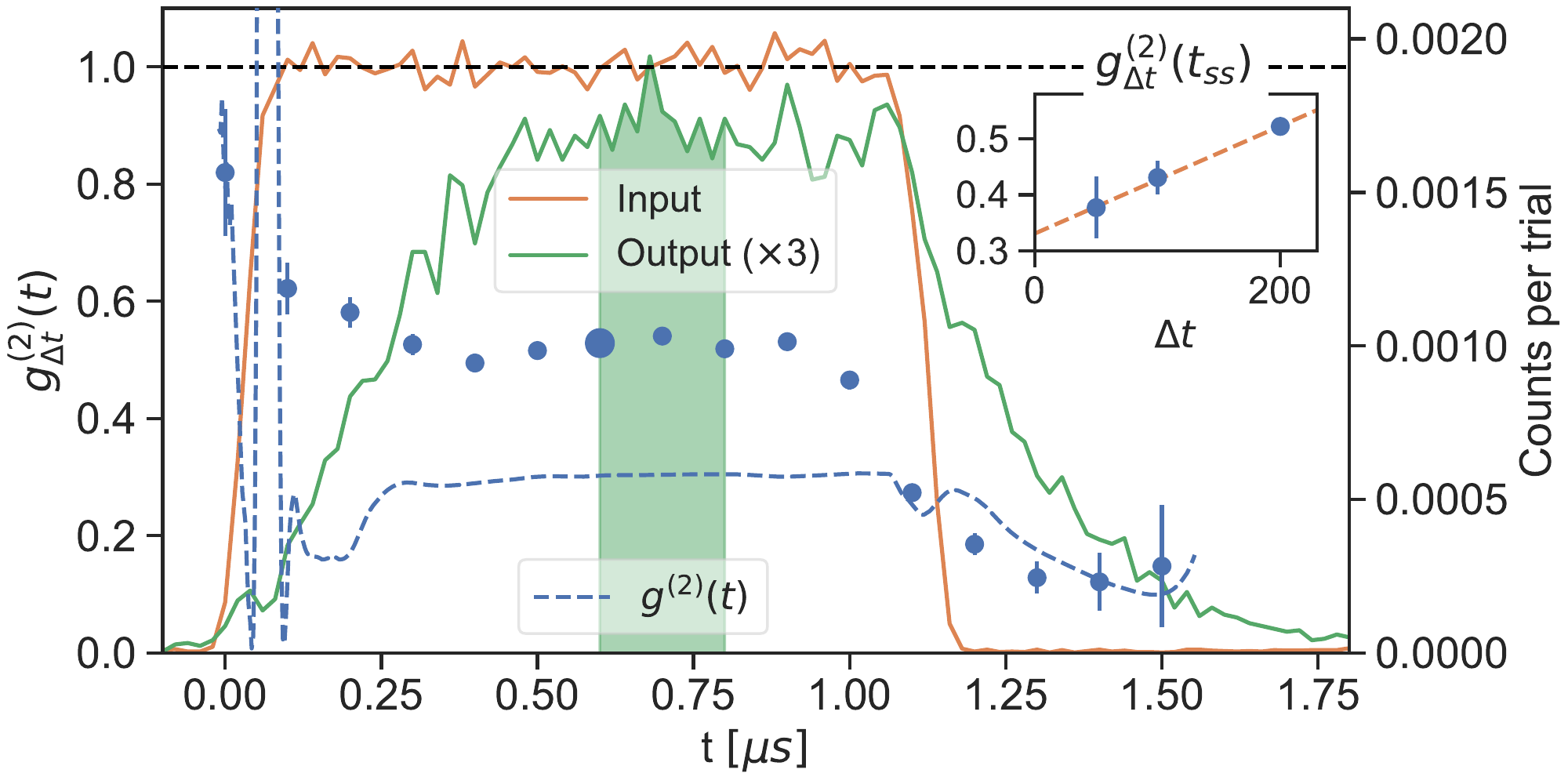}
	\caption{Transients for a square pulse. The counts per trial (right y-axis) of the input (orange) and output (green) pulses are shown with respect to the arrival time to the photodetector. An example of the 200 ns window that we take to calculate the $g^{(2)}_{\Delta t}(t)$ is shadowed. The measured second-order autocorrelation function (blue points, left y-axis) is shown for different starting times $t$ of the 200 ns window along the output pulse. The highlighted point corresponds to the shadowed region. The dashed blue line shows the second-order correlation function obtained from the theoretical simulation of $g^{(2)}(t)$ presented in Sec. \ref{theory}, which corresponds to the case with $\Delta t \to 0$. For comparison, in the inset plot, we show that the average steady-state (ss) experimental value $g^{(2)}_{\Delta t}(t_{ss})$ tends to the theoretical $g^{(2)}_{ss}(t) \approx0.3$ for smaller $\Delta t$ (the dashed orange line is a linear fit to guide the eye). Note that using a $\Delta t >0$ would smooth out the discontinuities seen in the simulation of the turn-on transient and could explain why they are not seen in the experimental data. The error bars correspond to one standard deviation.}
	\label{fig:square}
\end{figure}

To get a better understanding of the dynamics of Rydberg polaritons, we perform a more detailed study of the second-order autocorrelation function within a pulse. To this end, we select a detection window $\Delta t$ much shorter than the pulse duration and we measure the zero-delay autocorrelation function $g^{(2)}_{\Delta t}(t,0)$ as a function of the starting time of the detection window $t$. For simplicity, $g^{(2)}_{\Delta t}(t,0)\equiv g^{(2)}_{\Delta t}(t)$ henceforth. An example is shown in Fig. \ref{fig:square} where we show the $g^{(2)}_{\Delta t}(t)$ (blue points), obtained by taking a window of $\Delta t=200$ ns (green shadowed region), for different positions of this window throughout the output pulse. 

We find that the second-order autocorrelation function is not constant throughout the pulse duration, but decreases towards the end of the pulse. We observe three different regimes. For early times ($t<0.4 \ \mu$s), a first transient is observed, exhibited by a decrease in the $g^{(2)}_{\Delta t}(t)$ and an increase in the output intensity over time. We measure $g^{(2)}_{\Delta t}(t)=0.8 \pm 0.1$ for $t=0$ and $g^{(2)}_{\Delta t}(t)=0.53 \pm 0.02$ for $t= 0.3\ \mu$s. In the steady-state region ($0.4 \ \mu$s $<t<1 \ \mu$s), both the $g^{(2)}_{\Delta t}(t)$ and the output intensity are constant over time. Note that in this region, the value of $g^{(2)}_{\Delta t}(t)$ decreases for lower $\Delta t$, as shown in the inset of Fig. \ref{fig:square}, since the probability of a two-photon event decreases for lower delay times $\tau$ (see Fig. \ref{fig:g2ex}.b). When the input pulse is switched off ($t>1 \ \mu$s), a second transient is observed. Although a decrease in the output intensity is expected, $g^{(2)}_{\Delta t}(t)$ follows the same behavior in the same time range, decreasing until reaching a value as low as $g^{(2)}_{\Delta t}(t)=0.12 \pm 0.05$ for $t=1.4 \ \mu$s. These changes of $g^{(2)}_{\Delta t}(t)$ along the pulse correspond to three distinguishable processes, happening at different times during the propagation of Rydberg polaritons, as we will describe in the sec. \ref{theory}.
 
\begin{figure}[h]
	\centering
	\includegraphics[width=0.9\textwidth]{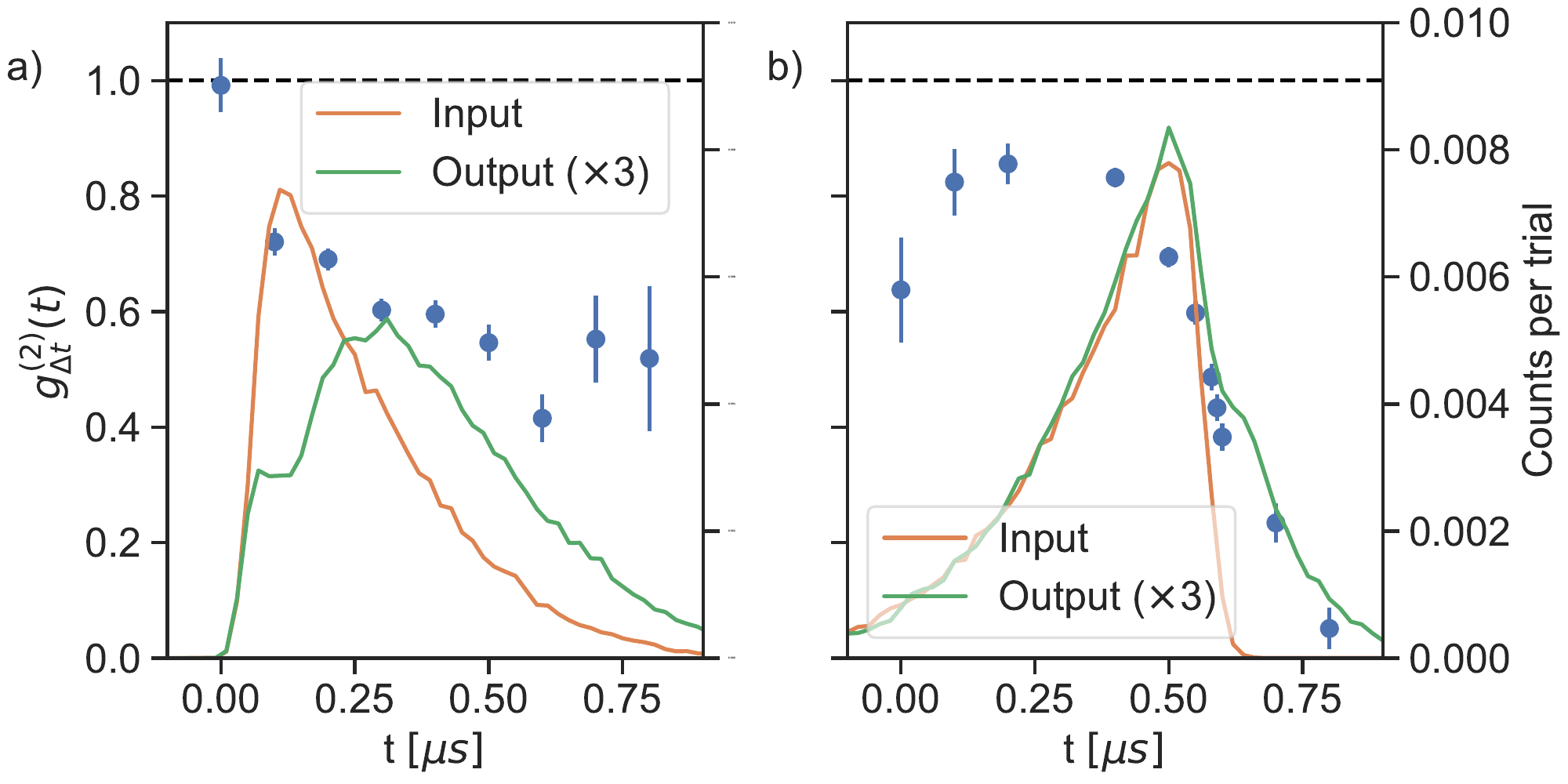}
	\caption{Transients for triangular pulses. Second-order autocorrelation function (in blue) for a window $\Delta t$ of 200 ns along the output pulse, temporal distribution of the input (in orange) and output (in green) pulses for an input pulse with a pseudo-triangular shape and a negative slope (a) and for an input pulse with a positive slope (b). The error bars correspond to one standard deviation.}
	\label{fig:triangular}
\end{figure}

To show the dependence of these transients with the shape of the input pulse, we study the propagation of pseudo-triangular pulses (see Fig. \ref{fig:triangular}), following the same method described above. In the case of a triangular shape with a negative slope (see Fig. \ref{fig:triangular}.a), the probe field is switched on abruptly but slowly switched off. Here, we observe that $g^{(2)}_{\Delta t}(t)$ starts with a value close to 1, but then it decreases rapidly towards smaller values, remaining constant at the end of the pulse with $g^{(2)}_{\Delta t}(t)=0.52\pm0.13$ for the last point ($t= 0.8 \ \mu$s). For a triangular shape with a positive slope (see Fig. \ref{fig:triangular}.b), i.e. slowly switched on and abruptly turned off, we only observe a clear transient at the end of the pulse, since $g^{(2)}_{\Delta t}(t)$ starts to decrease when the input pulse intensity goes to zero ($t>1 \ \mu$s). A value of $g^{(2)}_{\Delta t}(t)=0.05\pm0.04$ is obtained for the last point ($t= 0.8 \ \mu$s), which is much lower than the observed value for the triangular shape with negative slope.

These results show that the appearance of the transients depends on how the input pulse varies over time. Specifically, very low values of $g^{(2)}_{\Delta t}(t)$ are observed at the end of the pulse only if the decrease in intensity of the input pulse is fast enough.

\section{Theoretical model}\label{theory}
Here, we numerically analyze the transient behavior observed experimentally and also provide an intuitive model that elucidates the underlying physics in the simplest case of a fully blockaded medium. Afterward, we use this model to understand the experimental results. We first briefly discuss our numerical ``spin model" technique. The features of the output field are expected to only depend on the control field $\Omega_c$, on the optical depth $D$ of the entire medium and on the optical depth $D_b$ per blockade radius, but not on the total number of atoms or on the per-atom coupling efficiency of the probe mode separately. Taking advantage of this observation, it is then possible to investigate an artificial, quasi-one-dimensional system of a much smaller, tractable number of atoms with increased coupling efficiency to the probe, while maintaining the same $D$, $D_b$ of the experiment. In all subsequent numerical simulations, we fix the optical depth per atom on the $\ket{g}$-$\ket{e}$ transition (absent EIT) to be $D_{atom}\approx 0.36$ and vary the number of atoms to tune $D$. Furthermore, the dynamics of the atomic internal degrees of freedom (``spins’’) are encoded in an interacting spin model, whose solution can be used to re-construct all field properties via an input-output formalism. Details of this formalism can be found in Refs. \cite{Caneva2015,Manzoni2017} and its previous application toward modeling Rydberg EIT experiments is described in Ref. \cite{Bienias2020}. Our numerical calculation truncates the Hilbert space to two total atomic excitations (in the states $\ket{e}$ and/or $\ket{r}$) and ignores quantum jumps, which is valid in the regime of weak probe light \cite{Caneva2015}.  Moreover, we will restrict ourselves to the case where the probe field and the control field are resonant with the $\ket{g}$-$\ket{e}$ and $\ket{e}$-$\ket{r}$ transitions, respectively.

\begin{figure}[!htb]
\centering%
{\includegraphics[width=1.0\textwidth]{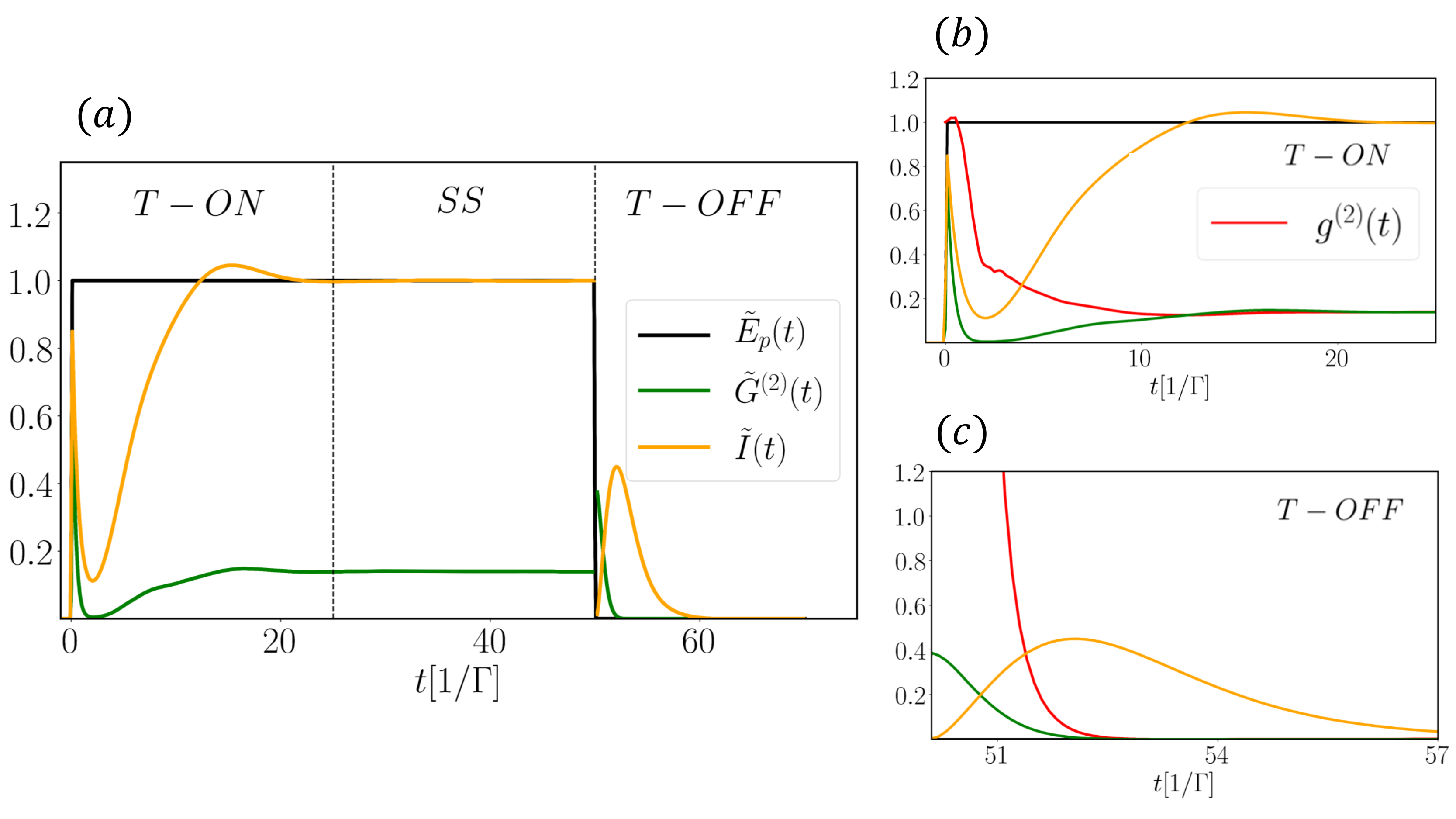}}\quad\quad
\caption{Analysis of the dynamics of a weak square input pulse for a control Rabi frequency of $\Omega_c=\Gamma/2$, $D\approx3.6$ and a fully blockaded system. (a) The normalized output intensity $\tilde{I}(t)$ as a function of time (yellow) and the correlation function $\tilde{G}^{(2)}(t)$ (green). The intensity of the input pulse $\tilde{E}_p$ (normalized to unity) is also shown in black for reference. Three distinct regions in the dynamics can be observed, turn-on (T-ON), steady state (SS) and turn-off (T-OFF). (b) Normalized second-order correlation function $g^{(2)}(t)$ (red), output intensity $\tilde{I}(t)$ (yellow) and $\tilde{G}^{(2)}(t)$ (green) in the turn-on stage, with the input again shown for reference (black). (c) $g^{(2)}(t)$ (red), $\tilde{I}(t)$ (yellow) and $\tilde{G}^{(2)}(t)$ (green) in the turn-off stage.\label{SquarePulse}}
\end{figure}

Analytically, we consider the simplest model where similar transient behavior can be observed, consisting of the limit where the probe input field $E_p(t)$ approaches a square pulse with amplitude $E_{p_0}$ when turned on and where the system is fully blockaded, so that two atoms cannot simultaneously be in the Rydberg state $\ket{r}$. We consider a square pulse long enough that all observables equilibrate to a steady-state value at some point during the period where the pulse is turned on, and we assume that the dephasing of the Rydberg state is negligible. In Fig. \ref{SquarePulse} we show a representative plot versus time of the normalized output intensity $\tilde{I}(t)=\bra{\psi}E^{\dagger}(t)E(t)\ket{\psi}/E_{p_0}^2$, the normalized ``two-photon intensity'' $\tilde{G}^{(2)}=\bra{\psi}E^{\dagger 2}(t)E^2(t)\ket{\psi}/E_{p_0}^4$ and the normalized second-order correlation function at zero delay $g^{(2)}(t,0)=\tilde{G}^{(2)}(t)/\tilde{I}^2(t)\equiv g^{(2)}(t)$. We can clearly identify three separate regimes of behavior: the initial turn-on, the steady state and the final turn-off. Below, we analyze each regime separately in more detail.

\begin{figure}[!htb]
\centering%
{\includegraphics[width=0.85\textwidth]{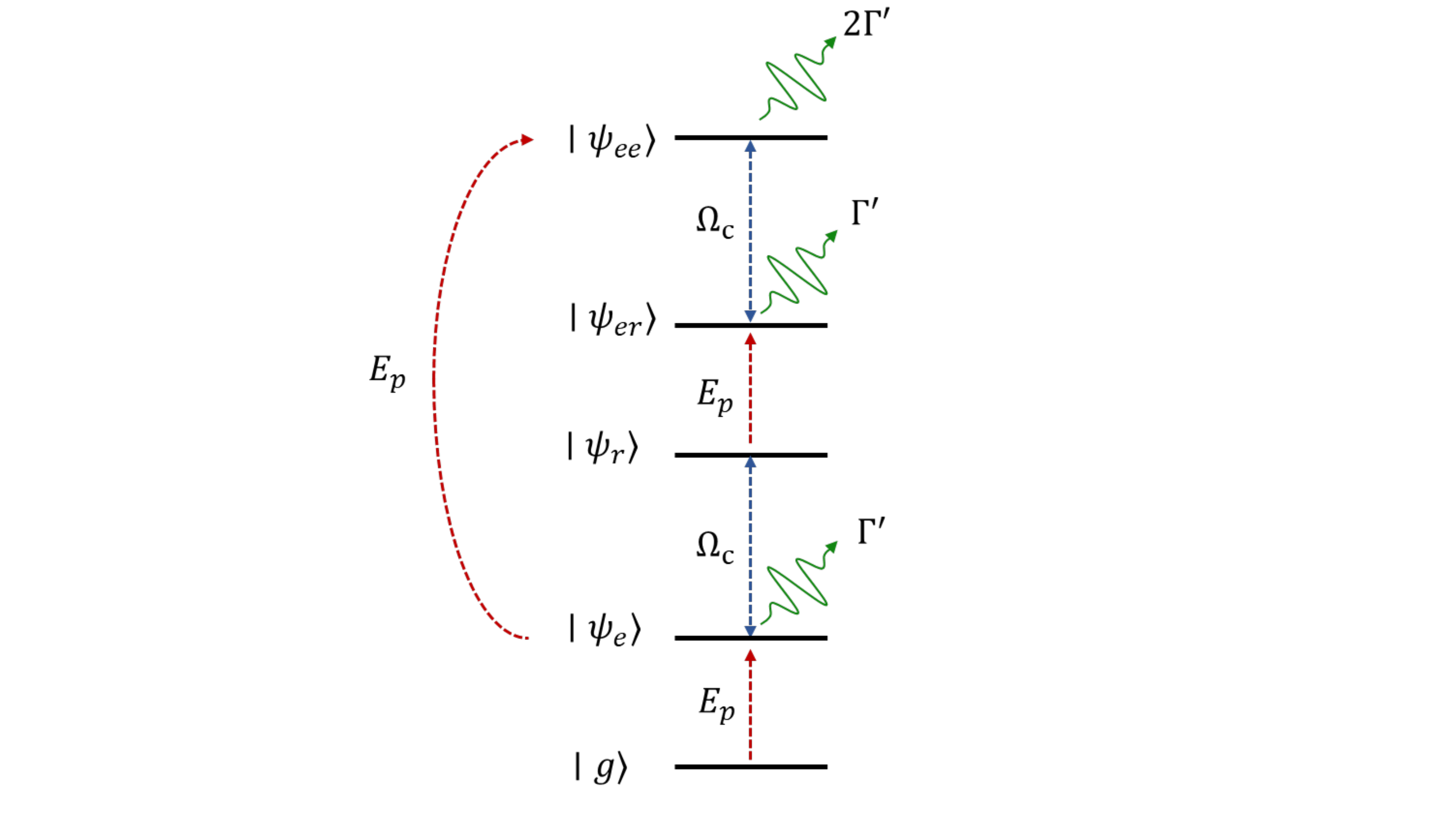}}\qquad\qquad
\caption{Schematic representation of the atomic level scheme for a fully blockaded medium in the weak driving regime. Within the single-excitation manifold, one atom can either be in the excited state $\ket{e}$ or in the Rydberg state $\ket{r}$, as represented by the subspaces $\ket{\psi_{e,r}}$, respectively. Similarly, in the two-excitation manifold, either one atom can occupy the state $\ket{e}$ while another one the state $\ket{r}$, or two atoms can simultaneously occupy the state $\ket{e}$. The pathways by which the subspaces can become excited by either the probe ($E_p$) or by the control field ($\Omega_c$) are indicated by dashed arrows, while the free-space emission rates due to decay of the excited state $\ket{e}$ by green wavy arrows. \label{Level_Structure}}
\end{figure}

%%%%%%%%%%%%%%%%%%
\subsection{Steady state}
%%%%%%%%%%%%%%%%%%
We begin with the steady-state properties, which have already been extensively studied in Ref. \cite{Peyronel2012}. First, we note that in the weak driving regime, the output intensity is predominantly dictated by the single-excitation component of the atomic system (since the one- and two-photon populations of the incoming pulse scale like $E_p^2$ and $E_p^4$, respectively). As the single-photon component does not experience Rydberg nonlinearities and sees perfect transparency associated with EIT, it results that the normalized intensity $\tilde{I}(t)\approx1$ in the steady state (yellow curve in Fig.~\ref{SquarePulse}.a). On the other hand, the two-photon intensity $\tilde{G}^{(2)}$ is efficiently attenuated due to the destruction of the EIT transparency condition by the Rydberg blockade. For a fully blockaded medium, the steady-state value of the output second-order correlation function was proved to be decreasing with $D$ and increasing with $\Omega_c$ and to admit the following approximate expression in the limit of high optical depth: $g^{(2)}_{ss}\approx 4\frac{1+\left(\Omega_c/\Gamma\right)^2}{\pi D}\exp\left[\frac{-D}{1+\left(\Omega_c/\Gamma\right)^2}\right]$ \cite{Peyronel2012}.

%%%%%%%%%%%%%%%%%
\subsection{Turn-on}
%%%%%%%%%%%%%%%%%%
Once we switch the probe field suddenly on at $t=0$, the system evolves toward the steady-state configuration. In Fig.~\ref{SquarePulse}.b, we show a representative plot of $g^{(2)}(t)$ during this transient period. To further understand the behavior, it is convenient to utilize an input-output relation, which formally allows one to express the output field operator as a sum of the input field and the field re-emitted by the atoms. In a quasi-one-dimensional propagation problem, this takes the form \cite{Caneva2015,Manzoni2017}:

\begin{equation}\label{ElectricField}
	E=E_p(t)-i\sqrt{\frac{\Gamma_{1D}}{2}}\sum_{h=1}^N e^{ik_pz_h}\sigma_{ge}^h.
\end{equation} 
Here, $\sigma_{ge}^{h}=\ket{g}\bra{e_h}$ is the $h$-th atomic lowering operator, where $\ket{g}$ is the ground state of the system and $\ket{e_h}$ states for atom $h$ in the $\ket{e}$ state and all the others in the ground state, $z_h$ is the position of atom $h$ and $k_p\approx \omega_{eg}/c$ is the probe beam wavevector, where $c$ is the speed of light in vacuum. We have normalized the field such that $E^{\dagger}E$ has units of photon number per unit time and $\Gamma_{1D}$ is the emission rate of atoms into the Gaussian mode defined by the input probe beam. The spontaneous decay rate $\Gamma$ of the excited state is decomposed into the sum of $\Gamma_{1D}$ and $\Gamma'$, which is the decay into noncollectable directions and represents the losses. The optical depth reads $D=2N\mathrm{log}\left(\frac{\Gamma_{1D}+\Gamma'}{\Gamma'}\right)$, and in our spin model simulations we fix $\Gamma_{1D}/\Gamma'=0.2$, so that the optical depth per atom results to be $D_{atom}\approx0.36$. Since the atoms are initially in the ground state and the atomic properties must evolve continuously, one finds $\sigma_{ge}^h\ket{\psi(t=0^+)}=0$ at a time $t=0^+$ immediately after the turn-on of the pulse. From Eq. \ref{ElectricField}, this implies that the output field is the same as the input field immediately after turn-on, and in particular, $g^{(2)}(0^+)=1$ reflects the coherent-state statistics of the input field.
This also causes the first intensity peak immediately after turn-on, shown in Fig. \ref{SquarePulse}.b, which can alternatively be thought as input light components traveling through the ensemble too fast to interact with the atoms.

We now characterize the time scale $\tau_0$ over which $g^{(2)}(t)$ is expected to approach its steady-state value, $g^{(2)}(\tau_0)\sim g^{(2)}_{ss}$. To do so, it is helpful to draw a schematic of the possible atomic levels that can be excited in the weak-probe limit (up to two excitations), as indicated in Fig.~\ref{Level_Structure}. Here, we denote $\ket{\psi_{e,r}}$ as the manifold of states where only one atom is excited to states $\ket{e},\ket{r}$, respectively. Similarly, $\ket{\psi_{er,ee}}$ denote the manifold of states where one atom is in $\ket{e}$ while another is in $\ket{r}$, or two atoms are in the state $\ket{e}$~(recall that we consider a fully blockaded medium, so two atoms cannot occupy state $\ket{r}$). We denote with dashed arrows the possible paths by which these states can be excited by the probe and control fields. We have also indicated by the wavy green arrows the rates of dissipation of these states due to spontaneous emission into $4\pi$.

In the limit of a weak probe beam, the population of the two-excitation manifold is sufficiently small that its back-action onto the evolution of the single-excitation manifold can be neglected. Furthermore, since the coupling of the probe and control fields between the ground state and the single-excitation manifold simply behave under EIT in the linear optics regime, their dynamics are straightforward to analyze. In particular, the square pulse should propagate through the system at the reduced EIT group velocity $v_{gr}$ and its leading-edge should reach the end of the medium in a time $\tau_{EIT}=L/v_{gr}=4D\Gamma'/\Omega_c^2$. The single-excitation manifold effectively acts as a source to populate the two-excitation one (see Fig. \ref{Level_Structure}), and the two-excitation manifold should reach the steady state within a time $\sim\Gamma^{-1}$ of the single-excitation manifold doing so, due to its natural dissipation. As this time is negligible, we then expect for $\tau_0\approx \tau_{EIT}$. 

\begin{figure}[!htb]
\centering%
{\includegraphics[width=0.9\textwidth]{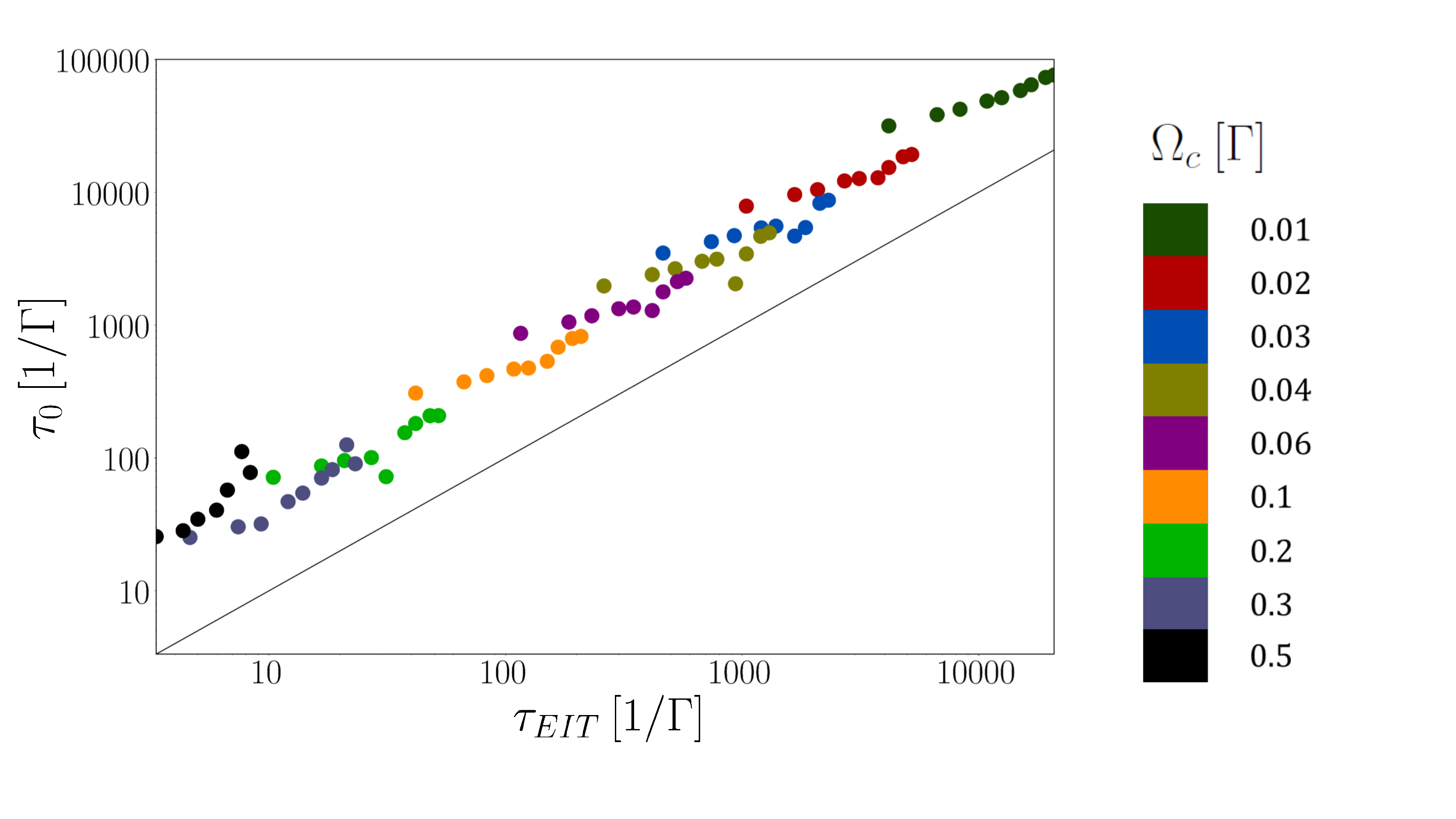}}\quad\quad
\caption{Relationship between the turn-on transient time $\tau_0$ and the EIT propagation time $\tau_{EIT}$ (black solid line). $\tau_0$ is defined as the time after which $\left|g^{(2)}(t)-g^{(2)}_{ss}\right|/g^{(2)}_{ss}<0.005$. Different values of $\tau_{EIT}=4D\Gamma'/\Omega_c^2$ are sampled by varying the optical depth $D$ from $1.8$ to $9.1$ and the control field $\Omega_c$ from $\Gamma/100$ to $\Gamma/2$. The values of $\Omega_c$ are indicated by the different colors, as represented in the colorbar to the right.  \label{T0scalingTeit}}
\end{figure}

In Fig. \ref{T0scalingTeit}, we confirm this scaling numerically. In particular, we vary both the control field amplitude $\Omega_c$ and optical depth $D$ over a large range of values. We further define $\tau_0$ as the time in which the numerically obtained fractional difference between the transient value of the second-order correlation function and its steady-state value $|g^{(2)}(\tau_0)-g_{ss}^{(2)}|/g_{ss}^{(2)}$ drops below $0.005$. We see that $\tau_0\approx\tau_{EIT}$ over the entire range of parameters studied, and independent of the specific values of $D$ and $\Omega_c$ separately. This is in agreement with the experimental data, where the duration of the first transient is $\sim400$ ns and we calculate $\tau_{EIT}\sim$ 450 ns, similar to the time the output intensity takes to arrive at its steady-state value (see Fig. \ref{fig:square}).

%%%%%%%%%%%%%%%%%%
\subsection{Turn-off}
%%%%%%%%%%%%%%%%%%
In Fig. \ref{SquarePulse}.c we show the output $\tilde{I}$, $\tilde{G}^{(2)}$ and $g^{(2)}$ when we switch the probe field suddenly off, starting from a steady-state initial condition. The dynamics exhibits two notable features. First, discontinuities can develop in the observables immediately after the shutoff. Second, one sees that the second-order correlation function $g^{(2)}(t)$ approaches zero at sufficiently long times, indicating a stronger antibunching than the one realizable in the steady state. We begin by analyzing the dynamics in the vicinity of the shutoff of the probe, which we define to occur at the time $\bar{t}$.

Using the input-output relation of Eq.~(\ref{ElectricField}), one sees that the output field operator $E$ evolves discontinuously. In particular, the outgoing intensity at a time $\bar{t}^+$ immediately following the shutoff, $\avg{E^\dagger (\bar{t}^+)E(\bar{t}^+)}$, will only be due to purely atomic emission and jump from its value at $\bar{t}^-$ immediately before the shutoff. Furthermore, since this intensity is dominated by the single-excitation component for weak driving, and as the $\ket{e}$-state component of this manifold is unpopulated ($\ket{\psi_e}=0$) due to perfect EIT, the atoms are not able to emit light instantaneously and one finds $\tilde{I}(\bar{t}^+)\approx 0$. Conversely, the normalized two-photon intensity $\tilde{G}^{(2)}$ experiences a discontinuous increase. 
In steady-state conditions and high optical depth, its value $\tilde{G}^{(2)}_{ss}\ll 1$ can be understood from Eq. \ref{ElectricField} as nearly perfect destructive interference between the incoming field and the field re-emitted by the atoms, since two photons cannot be efficiently transmitted due to the Rydberg blockade. Therefore, if the input field is instantly extinguished, the two-photon outgoing intensity is due to a purely atomic emission which, for large $D$,
is almost equal in amplitude (but opposite in phase) to the incoming field. As a consequence, one expects that $\tilde{G}^{(2)}(\bar{t}^+)\rightarrow 1$ and $g^{(2)}(\bar{t}^+)\rightarrow\infty$ in the square pulse limit. For a continuous switch off, a flash of bunched output light can still emerge if the time scale of the shutoff is faster than the time needed by the atoms to react, which is roughly $\sim\Gamma^{-1}$ \cite{Moehl2020}. 

\begin{figure}[!htb]
\centering%
{\includegraphics[scale=0.48]{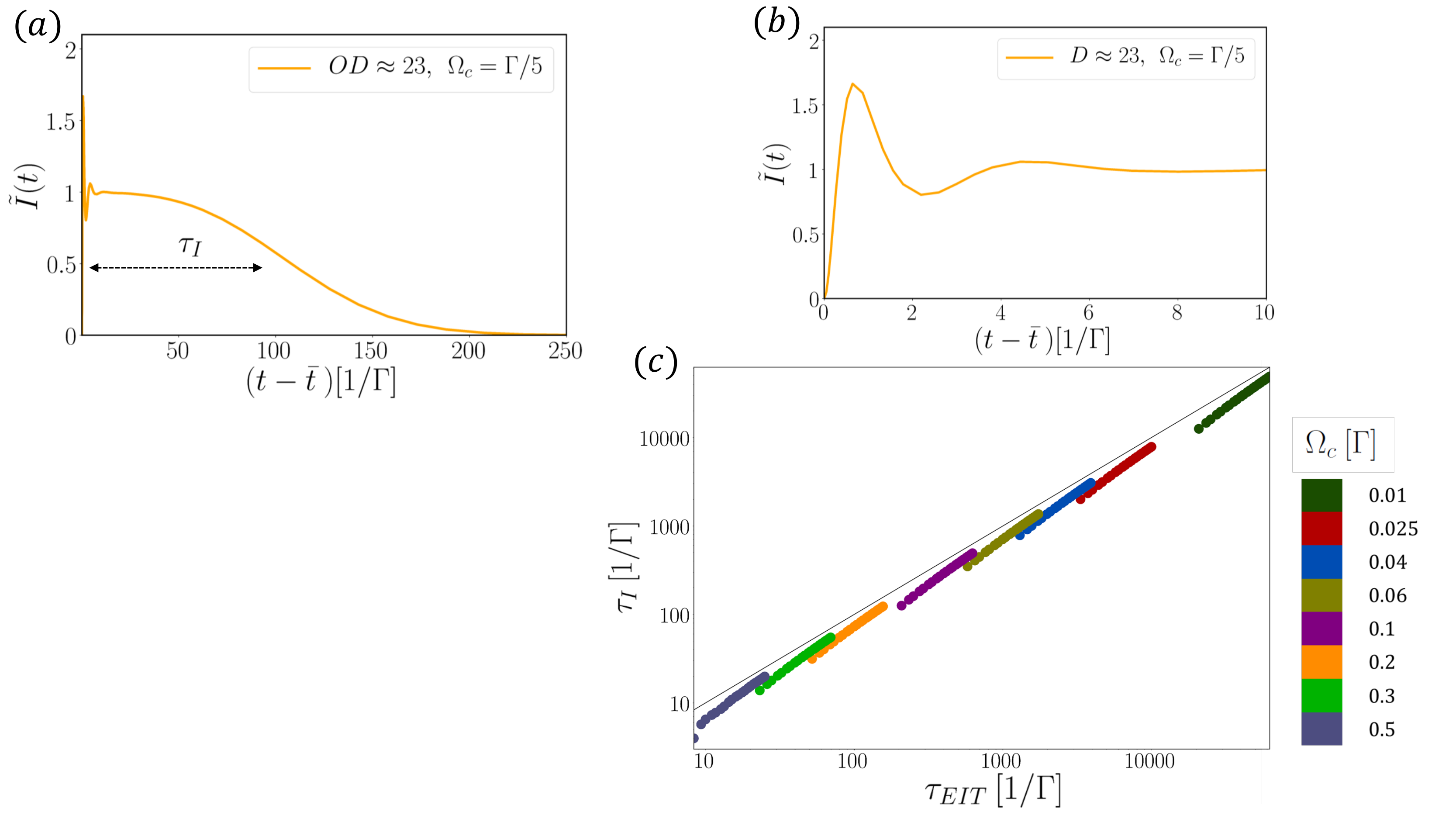}}\quad\quad
\caption{Analysis of the single-excitation dynamics in the turn-off regime. (a) Representative plot of the normalized output intensity $\tilde{I}(t)$ for $D\approx23$ and $\Omega_c=\Gamma/5$.  The single-excitation characteristic time $\tau_I$ is defined as the amount of time needed, following the turn-off (occurring at time $\bar{t}\:$), for the normalized intensity to drop to half of its steady-state value, $\tilde{I}(\bar{t}+\tau_I)=0.5$. (b) Detail of the short times dynamics. (c) Relationship between $\tau_I$ and the EIT propagation time $\tau_{EIT}$ sampled by varying $D$ form $9.1$ to $27.3$ and $\Omega_c$ from $\Gamma/100$ to $\Gamma/2$. The values of $\Omega_c$ are indicated by the different colors, as represented in the colorbar to the right. \label{Turn_Off_Single_Excitation}}
\end{figure}

The analysis of the transient dynamics following the instant shutoff of the probe is made easier by noting that the single- and the two-excitation components of the atomic system become decoupled when $E_p=0$ (see Fig. \ref{Level_Structure}) and we can thus investigate the dynamics in each manifold separately. Immediately following the shutoff, any atomic excitation in the single-excitation manifold takes the form of a spin wave in the Rydberg state $\ket{\psi_1(\bar{t}^+)}=\ket{\psi_1^{ss}}\sim\sum_he^{ik_pz_h}\ket{r_h}$ due to the perfect EIT condition. This makes the subsequent dynamics equivalent to the retrieval of a stored spin wave in an EIT-based quantum memory. We will restrict ourselves to the regime of reasonably high $D$ and small $\Omega_c$, such that the retrieved field takes a similar spatial form as the spin wave itself (in contrast when $\Omega_c
%\gtrsim
\geq
\Gamma$, the outgoing pulse can oscillate due to Rabi flopping dynamics between $\ket{e}$ and $\ket{r}$). In Fig. \ref{Turn_Off_Single_Excitation}.a, we show a representative plot of the output intensity as a function of time following the shutoff, for $D\approx23$ and $\Omega_c/\Gamma=0.2$. Due to the finite bandwidth of the EIT transparency window, the shape of the retrieved output intensity is smoother than the flat rectangular shape of the steady-state spin wave itself. We define a characteristic time $\tau_I$ as the amount of time following the shutoff for the normalized outgoing intensity to reach half of its steady-state value,  $\tilde{I}(\bar{t}+\tau_I)=0.5$. We expect that $\tau_I\approx \tau_{EIT}$, i.e. $\tau_I$ should approximately coincide with the time needed to propagate across the medium at the reduced EIT group velocity. In Fig. \ref{Turn_Off_Single_Excitation}.c, we plot the numerically extracted $\tau_I$ over a broad range of optical depths and control field amplitudes and see a good agreement with the expected result. For the experimental data, the calculated value of $\tau_{EIT}$ is also in agreement with the decay time of the second transient (see Fig. \ref{fig:square}). Finally, we remark that at short times following the shutoff, the outgoing intensity can exceed the steady-state intensity of the original square wave input, $\tilde{I}(t)>1$, as illustrated in Fig. \ref{Turn_Off_Single_Excitation}.b. Physically, at short times, the control field can drive the original spin wave stored in the state $\ket{r}$ into a ``bright'' spin wave $\sim \sum_he^{ik_pz_h}\ket{e_h}$. It is known that such a spin wave experiences collectively enhanced or superradiant emission into the forward direction \cite{Bromley2016,Svidzinsky2010}, at a rate $\sim \Gamma D/4$, which physically arises from the constructive interference between the light emitted by different atoms in this direction due to their relative phases $\sim e^{ik_pz_h}$.

\begin{figure}[!htb]
\centering%
{\includegraphics[scale=0.5]{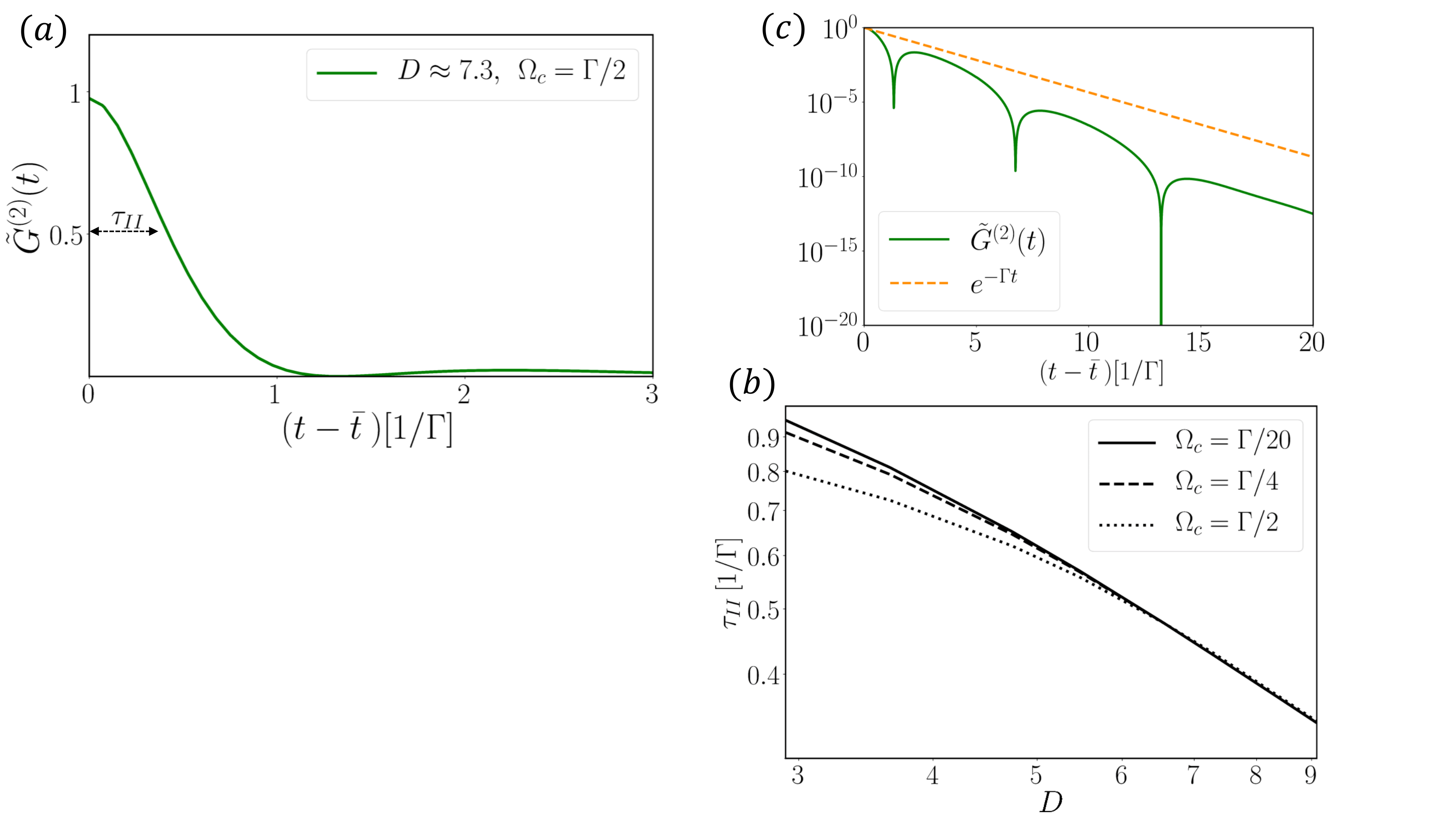}}\quad\quad
\caption{Analysis of the double-excitation dynamics in the turn-off regime. (a) Representative plot of the normalized output two-photon intensity $\tilde{G}^{(2)}(t)$ for $D\approx7.3$ and $\Omega_c=\Gamma/2$. The double-excitation characteristic time $\tau_{II}$ is defined as the amount of time needed, following the turn-off (occurring at time $\bar{t}\:$), for the normalized two-photon intensity to drop to half value immediately after turn-off, $\tilde{G}^{(2)}(\bar{t}_{II})=\tilde{G}^{(2)}(\bar{t})/2$. (b) The double-excitation characteristic time $\tau_{II}$ sampled by varying $D$ from $2.9$ to $9.1$ for $\Omega_c=\Gamma/20,\:\Gamma/4,$ and $\Gamma/2$. (c) Long time profile of $\tilde{G}^{(2)}(t)$ and reference spontaneous decay $e^{-\Gamma t}$ (dashed yellow line). \label{Turn_Off_Double_Excitation}}
\end{figure}

We now turn to the two-photon intensity. In Fig. \ref{Turn_Off_Double_Excitation}.a we show a representative plot of the output $\tilde{G}^{(2)}(t)$ at the turn-off. One sees that at short times it rapidly decays to zero with a hump-like profile, while at long times it exhibits slower oscillating features. We define a characteristic temporal length of the short time behavior, $\tau_{II}$, as the amount of time for the normalized two-photon intensity to reach half of its value immediately after turn-off, $\tilde{G}^{(2)}(\bar{t}+\tau_{II})=\tilde{G}^{(2)}(\bar{t})/2$. Numerically, we find that $\tau_{II}$ scales like $1/D$, as shown in Fig. \ref{Turn_Off_Double_Excitation}.b. This can be understood as both the $er$ and the $ee$ components of the two-excitation steady-state wave function $\ket{\psi_2^{ss}}$ are phase-matched and emit in a collectively enhanced fashion immediately after turn-off. However, as $\ket{\psi_2^{ss}}$ is not an eigenstate of the system, it can undergo spatio-temporal evolution. At long times, the emitted two-photon intensity will thus no longer be collectively enhanced but will decay at a rate $\sim\Gamma$ comparable to the single-atom decay rate. This behavior can be seen in Fig. \ref{Turn_Off_Double_Excitation}.c, by comparing the envelope of $\tilde{G}^{(2)}$ at long times with $e^{-\Gamma t}$. 

We can now finally understand why $g^{(2)}(t)=\tilde{G}^{(2)}(t)/\tilde{I}(t)^2$ decreases below its steady-state value at long times after the turn-off. In particular, with increasing optical depth, the numerator $\tilde{G}^{(2)}(t)$ describing the two-excitation component rapidly decays on a time scale $\tau_{II}\propto 1/D$ due to collective enhancement (with a small residual component decaying at the single-atom rate $\sim\Gamma$), while the denominator persists for a longer time $\tau_{I}\propto D$ due to the slow retrieval dynamics of the single-excitation component.
%%%%%%%%%%%%%%%%%%
\subsection{Simulation of the experiment}
%%%%%%%%%%%%%%%%%%
To reproduce the experimental results, we insert into our spin model the optical depth, the spontaneous decay rate and the control field values given in Sec. \ref{sec:Experiment}. We also add the motional dephasing $\gamma_r$ of the Rydberg atoms. The numerical results thus obtained are in good agreement with the experimental EIT transmission and with the output light pulse, shown in Fig. \ref{fig:setup}.d and \ref{fig:g2ex}.a, respectively. We remove the fully blockaded hypothesis and model the finite blockade with an interaction potential of the form $V_r=V_0\left(r_b/r\right)^6$, where $r_b$ is the blockade radius and $V_0=2\Omega_c^2\left[\Gamma_{1D}\left(2\Gamma'+\Gamma_{1D}\right)\right]^{-1/2}$ is the single-atom bandwidth. By setting an optical depth per blockade radius of $D_b\approx0.9$, one finds $g^{(2)}(t)\approx0.3$, which is in agreement with the $\tau=0$ value of Fig. \ref{fig:g2ex}.b. In Fig. \ref{fig:square}, the resultant pulse analysis is shown and it is worth noticing how the measured $g_{\Delta t}^{(2)}(t) $ tends towards the numerically predicted $g^{(2)}(t)$ as one decreases the length of the detection window $\Delta t$.

\section{Single-photon generation and storage}
\begin{figure}[h]
	\centering
\includegraphics[width=0.9\textwidth]{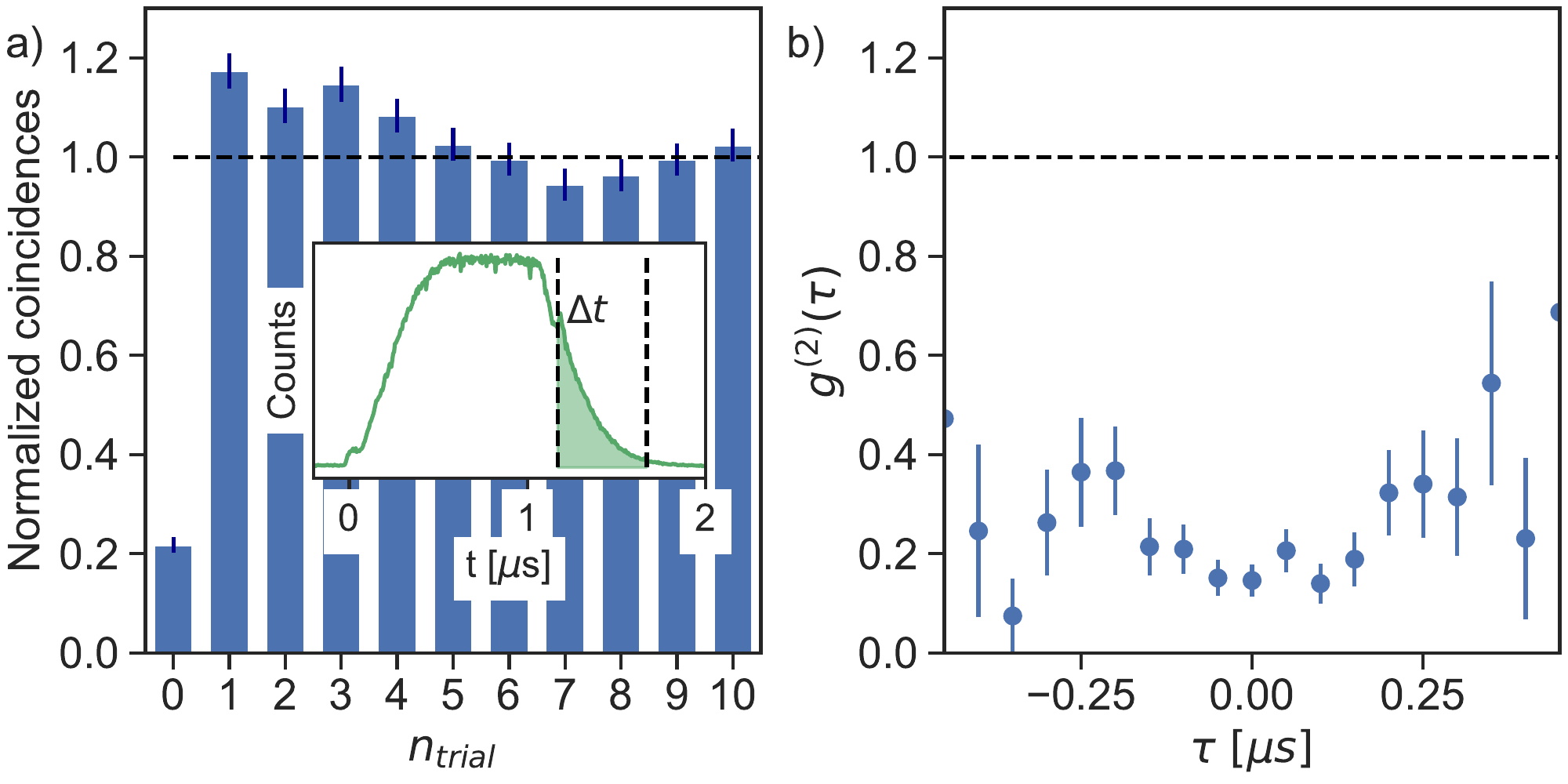}
	\caption{Single-photon generation: example of single-photon generation with  $\Delta t$ = 500 ns. (a) Normalized coincidence counts between different trials, leading to a value of $g^{(2)}_{\Delta t}= 0.218\pm0.015$ for $n_{trial}=0$. (b) Time-resolved measurement of $g^{(2)}(\tau)$ leading to a zero-delay value of $g^{(2)}(0)=0.15\pm0.03$. Error bars correspond to one standard deviation.}
	\label{fig:sp500}
\end{figure}

The turn-off transient, where the $g^{(2)}_{\Delta t}(t)$ value strongly decreases, opens the way for possible applications related to narrowband single-photon generation. For that purpose, we could cut the output pulses and exploit the single photons arriving in the last part of the pulse. To analyze this proposal, we select a temporal window at the end of the pulse where looking for detection counts and coincidences. Then, we measure the $g^{(2)}_{\Delta t}(t)$ values of the photons arriving inside this window and their corresponding detection and generation probability. The generation probability is inferred as $P_{g}=N_{1}/\eta_{det}$, where $N_1$ is the number of counts per trial arriving to SPAD$_1$ in the selected time window and $\eta_{det}$ is the detection efficiency, including fiber coupling, transmission through all the optical elements and single-photon detector efficiency. 
Fig. \ref{fig:sp500} shows an example of the single-photon generation for a time window $\Delta t$ = 500 ns and $t= 1.2 \ \mu$s (see inset plot). When measuring the normalized coincidences for consecutive pulses (see Fig. \ref{fig:sp500}.a), we get a value of $g^{(2)}_{\Delta t}=0.218\pm0.015$ significantly lower than for $\Delta t = 1.6 \mu s$. Moreover, the time resolved measurement within one pulse (see Fig. \ref{fig:sp500}.b) shows that the values of  $g^{(2)}(\tau)$ remain low for the full window, showing the generation of a localized single photon.

\begin{figure}[h]
	\centering
	\includegraphics[width=0.9\textwidth]{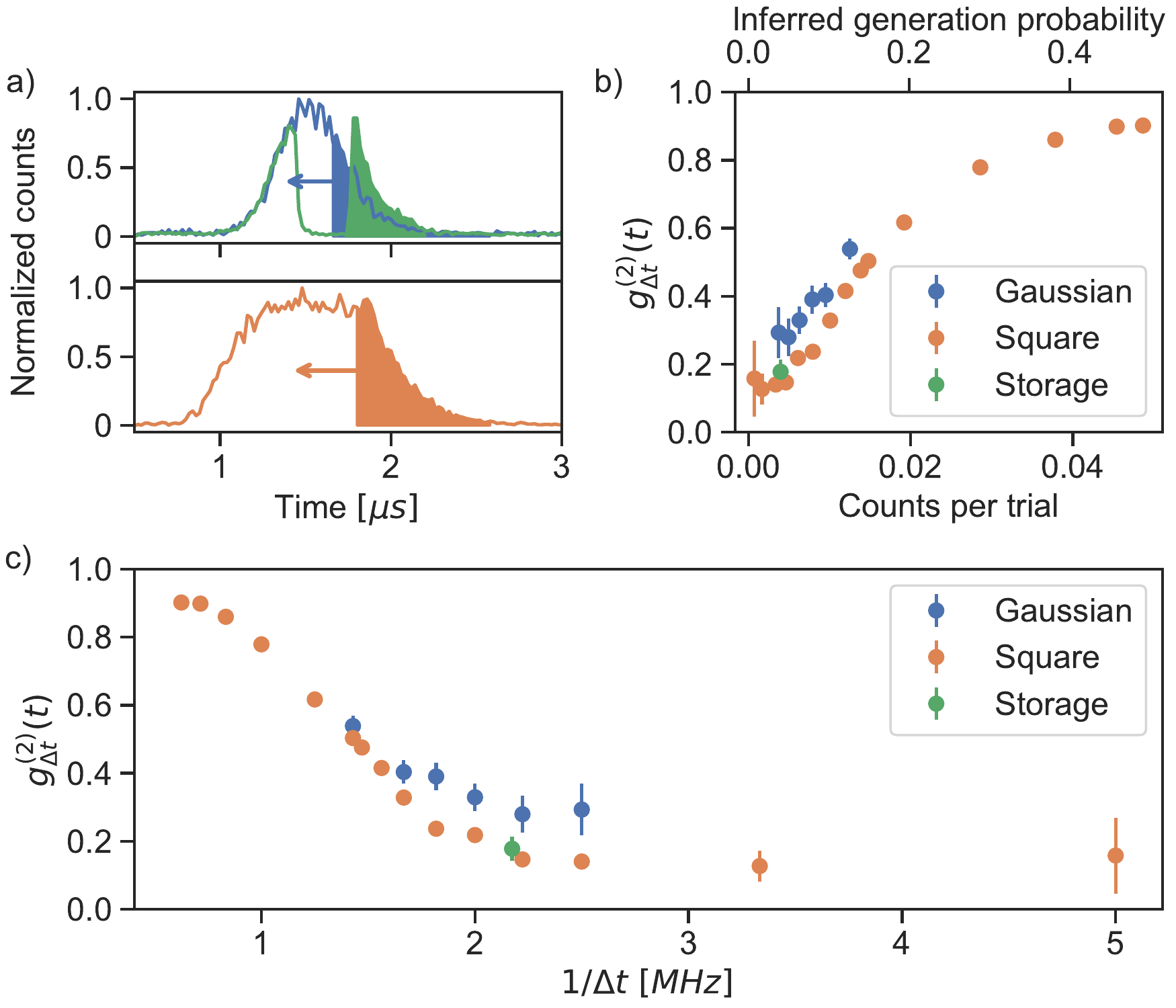}
	\caption{Single-photon generation. (a) Temporal distribution of a Gaussian pulse after propagation through EIT (in blue) and after storage (in green) and that of a square pulse after propagation through EIT (in orange). The temporal window is varied in plots (b) and (c) for the slow light case, as indicated by the arrow. (b) $g^{(2)}_{\Delta t}(t)$ as a function of the number of counts per trial in SPAD$_1$ (bottom axis) and the probability to have a photon at the output of the cloud (upper axis), for an input Gaussian pulse (in blue), the previously shown square pulse (in orange) and after storing for 500 ns (in green). The generation probability is increased by taking the same final time of the temporal window but changing the starting time to increase $\Delta t$, as we can see in plot (a). (c) $g^{(2)}_{\Delta t}(t)$ as a function of the inverse of the temporal window, for the same cases as before. Vertical error bars correspond to one standard deviation.}
	\label{fig:sp}
\end{figure}  

To further study the effect of the temporal window, we then vary $\Delta t$, while keeping the final time $t+\Delta t = 1.7 \ \mu$s fixed. Fig. \ref{fig:sp} shows the results for different duration of the temporal window, for Gaussian and square input pulses. When we increase the time window, the probability to have a detection count increases, but at the expense of reducing the quality of the single photons. For the previously studied square pulse, $g^{(2)}_{\Delta t}(t)=0.48 \pm 0.01$ is obtained for $\Delta t= 0.68 \ \mu$s and a generation probability of $0.145 \pm 0.014$, while $g^{(2)}_{\Delta t}(t)=0.147 \pm 0.017$ is obtained for $\Delta t= 0.45 \ \mu$s and a generation probability of $0.046 \pm 0.004$. We compare the results for the square pulse with the case of a Gaussian input pulse. We see that the Gaussian pulse leads to a higher value of $g^{(2)}_{\Delta t}(t)$, for all generation efficiencies, compared to the square pulse. This confirms that a fast shutoff of the input pulse is beneficial for the production of single photons.

We also compare the results with the strategy consisting of storing the input Gaussian pulse in the Rydberg state \cite{Maxwell2013, Distante2016, Distante2017, SchmidtEberle2020}. This can be made by switching off the control beam ($\Omega_c\rightarrow0$) when polaritons are traveling inside the medium. After a storage time of $t_S$ = 500 ns, we can retrieve the output pulse by switching on again the control beam. This method allows for an increase in the quantum character of the output pulse without changing the experimental conditions. Due to our limited $D$, the entire input pulse cannot be compressed inside the medium and only a part of the pulse can be stored as Rydberg excitations, importantly reducing the output efficiency.
In Fig. \ref{fig:sp}, we show the $g^{(2)}_{\Delta t}$ obtained after storage and retrieval (in green), for $\Delta t$ taking into account the whole retrieved pulse (see Fig. \ref{fig:sp}.a). As we can observe, the $g^{(2)}_{\Delta t}$ in the storage case is similar to the obtained in the turn-off transient of the transmitted square pulse, for the same generation probability (see green and orange points of Fig. \ref{fig:sp}.b). However, for lower quality of the output photons, i.e higher values of $g^{(2)}_{\Delta t}$, the generation probability in the transient case importantly increases.

Finally, in Fig. \ref{fig:sp}.c we plot $g^{(2)}_{\Delta t}$ as function of the inverse of $\Delta t$, which is proportional to the photon bandwidth. Values of $g^{(2)}(\Delta t)$ $<$ 0.2 can be achieved for $1/\Delta t\sim2$ MHz, showing that high quality, narrowband single photons can be generated with this technique. 

As shown in Sec. \ref{theory}, increasing the optical depth and reducing the Rabi frequency of the control field results in a better separation between the two-photon dynamics and the single-photon one and therefore leads to a smaller value of $g^{(2)}_{\Delta t}$ in the turn-off transient. However, the single-photon generation efficiency is currently affected by the low value of the transmission in the EIT transparency window, which is likely limited by the decoherence rate of the $|g\rangle$ to $|r\rangle$ transition, which also enforces a lower bound in the choice of $\Omega_c$. We expect that reducing the laser linewidths (currently around 300 kHz) by active stabilization on an optical cavity would bring to a significant increase in the EIT transmission and also make lower choices of the Rabi frequency of the control laser possible. Moreover, a larger value of $D_b$ would allow a more efficient compression of the pulse within a blockade radius. The current values of generation efficiency achieved (around 10 $\%$) are comparable with the values reported in single-photon generation experiments using off-resonant (Raman) excitation to the Rydberg states with Rydberg cold atomic ensembles \cite{Dudin2012}. It is informative to compare this efficiency with techniques of probabilistic single-photon generations with atomic ensembles, such as e.g. the Duan-Lukin-Cirac-Zoller scheme \cite{Duan2001}. This scheme generates probabilistically photon pairs in a two-mode squeezed state with a probability per trial $p$, where one of the photons is stored as a collective atomic spin excitation in the ensemble. Upon detection of the first photon which provides a heralding signal, this collective spin excitation can then be efficiently transferred into a single photon in a well-defined spatio-temporal mode with an efficiency $\eta_R$. The probability to generate a single photon per trial is therefore given by $P_{DLCZ}=p\eta_D\eta_R$ where $\eta_D$ is the probability to detect the first photon. For a perfect two-mode squeezed state and for $p\ll1$, the second-order autocorrelation of the retrieved photon is $g^{(2)}(0)$=4$p$. In the best-case scenario (i.e. with unity detection and read-out efficiency), a DLCZ source could generate a photon with $g^{(2)}(0)$=0.1 with a probability of $P_{DLCZ}=0.025$ per trial. In practice, with finite detection and read-out efficiencies, this value will be even lower. Therefore, even though the single-photon generation efficiency demonstrated in this paper is quite modest and could still be largely improved, it compare favorably to probabilistic schemes.

\section{Conclusions}

In this paper, we investigated the propagation of weak coherent pulses in a cold atomic ensemble in the regime of Rydberg electromagnetically induced transparency. We found experimentally that the second-order correlation function of the output pulse depends on the time throughout the pulse and strongly varying during the transients of the pulse. In particular, the value of  $g^{(2)}_{\Delta t}$ strongly decreases towards the end of the traveling pulse. Through a spin model analysis, we were able to quantitatively predict the measured pulse dynamics, both at the linear and nonlinear level, and to provide a qualitative explanation, in the test scenario of a perfectly square pulse propagating in a fully blockaded medium. Taking advantage of this behavior, we explored the possibility of using this effect to generate localized single photons and showed that it has better efficiency than a probabilistic DLCZ-like source.

\section{Acknowledgment}

This project received funding from the Government of Spain (PID2019-106850RB-I00; Severo Ochoa CEX2019-000910-S), Fundació Cellex, Fundació Mir-Puig, Generalitat de Catalunya (CERCA, AGAUR), Gordon and Betty Moore Foundation through Grant No. GBMF7446 to H. d. R. and from the European Union's Horizon 2020 research and innovation program under Grant Agreement No. 899275 (DAALI).

DEC acknowledges support from the European Union’s Horizon 2020 research and innovation programme, under FET-Open grant agreement No 899275 (DAALI), AEI Europa Excelencia program (EUR2020-112155, project ENHANCE), and Quantum Flagship project 820445 (QIA), MINECO Severo Ochoa program CEX2019-000910-S, Generalitat de Catalunya through the CERCA program, Fundació Privada Cellex, Fundació Mir-Puig, Plan Nacional Grant ALIQS (funded by MCIU, AEI, and FEDER), and Secretaria d'Universitats i Recerca del Departament d'Empresa i Coneixement de la Generalitat de Catalunya, co-funded by the European Union Regional Development Fund within the ERDF Operational Program of Catalunya (project QuantumCat, ref. 001-P-001644)

\section{References}

\bibliographystyle{iopart-num}
\bibliography{References}

\end{document}